\documentclass[preprint, 12pt,
showpacs, 
,amsmath,amssymb
]{revtex4}

\newcommand{\be}{\begin{equation}}
\newcommand{\ee}{\end{equation}} 
\newcommand{\bea}{\begin{eqnarray}} 
\usepackage{graphicx}
\newcommand{\eea}{\end{eqnarray}}
\usepackage{epsfig}
\usepackage{bm}

\newcommand{\A}{\mathbb{A}} 
\newcommand{\B}{\mathbb{B}}
\newcommand{\C}{\mathbb{C}}
\newcommand{\I}{\mathbb{I}}

\newcommand{\AT}{\mathbb{A}^{\footnotesize{\hbox{\scriptsize{T}}}}} 
\newcommand{\barAT}{\bar {\mathbb{A}}^{\footnotesize{\hbox{\scriptsize{T}}}}} 
\newcommand{\ATT}{\mathbb{A}^{\footnotesize{\hbox{\scriptsize{2T}}}}} 
 
\newcommand{\hAT}{\hat \A^{\footnotesize{\hbox{\scriptsize{T}}}}}

\newcommand{\tr}{{\hbox{Tr}}}

\begin{document}

\title{Velocity-Gradient Probability Distribution Functions in a Lagrangian Model of Turbulence}
\author{L. Moriconi$^1$, R.M. Pereira$^2$ and L.S. Grigorio$^3$, }
\affiliation{$^1$Instituto de F\'\i sica, Universidade Federal do Rio de Janeiro, \\
C.P. 68528, CEP: 21945-970, Rio de Janeiro, RJ, Brazil}
\affiliation{$^2$Divis\~{a}ão de Metrologia em Din\^{a}mica de Fluidos, Instituto Nacional de Metrologia, 
Normaliza\c{c}\~{a}o e Qualidade Industrial, Av. Nossa Senhora das Gra\c{c}as 50, Duque de Caxias, 
25250-020, Rio de Janeiro, Brazil}
\affiliation{$^3$Centro Federal de Educa\c c\~ao Tecnol\'ogica Celso Suckow da Fonseca, 28635-000, Nova Friburgo, Brazil}

\begin{abstract}
The Recent Fluid Deformation Closure (RFDC) model of lagrangian turbulence is recast in 
path-integral language within the framework of the Martin-Siggia-Rose functional formalism. 
In order to derive analytical expressions for the velocity-gradient probability distribution 
functions (vgPDFs), we carry out noise renormalization in the low-frequency regime and find 
approximate extrema for the Martin-Siggia-Rose effective action. We verify, with the help of 
Monte Carlo simulations, that the vgPDFs so obtained yield a close description of the 
single-point statistical features implied by the original RFDC stochastic differential 
equations.
\end{abstract}

\maketitle

\section{\leftline{Introduction}} 

It has been long known, since the seminal work of Batchelor and Townsend \cite{batch_town}, that
spatial derivatives of a turbulent velocity field do not behave as gaussian random variables. 
The current view on this still barely understood phenomenon is that the non-gaussian fluctuations of 
the velocity gradients -- the hallmark of turbulent intermittency -- are likely to be 
related to the existence of long-lived coherent structures and to deviations from the Kolmogorov 
``K41"  scaling, both important ingredients in the contemporary phenomenology of turbulence 
\cite{frisch,davidson}.

The main notorious difficulties with first-principle theories of intermittency stand on
(i) the inadequacy of perturbative expansions to deal with the coupled dynamics of vorticity 
and the rate-of-strain tensor at high Reynolds numbers and (ii) the fact that the closed equations 
for the time evolution of the velocity gradient tensor are non-local in the space variables.
Notwithstanding the strong coupling/non-local issues, it is actually possible 
to devise simplified fluid dynamical models that would capture relevant qualitative 
features of the intermittent fluctuations of the velocity gradient tensor
\cite{meneveau}. Here, a fundamental role is played by the lagrangian framework of fluid 
dynamics, since it leads in a natural way to reduced-dimensional systems, in the form of 
either ordinary or stochastic differential equations for the time evolution of the velocity 
gradient tensor \cite{viei,cant,chertkov_etal, chevi_menev}.

The aforementioned lagrangian models have been mostly investigated by means of numerical 
integrations of the associated differential equations, which can then be compared to well-established 
results of alternative direct numerical simulations. There is, however, a large room for the exploration 
of analytical tools in the study of lagrangian models of intermittency, a direction we pursue here, joining
other authors in this effort \cite{chertkov_etal,naso_etal}. We focus our attention on one particularly 
interesting stochastic model, the Recent Fluid Deformation Closure (RFDC) model \cite{chevi_menev}, and 
derive reasonable approximations for its velocity gradient probability distribution functions (vgPDFs). 
We put into practice standard statistical field-theoretical procedures for the computation of effective 
actions through vertex renormalization \cite{amit,zinn-justin}, which are carried out in the context of the 
Martin-Siggia-Rose functional formalism \cite{msr,janssen,dominicis,cardy}. Our approach -- essentially a 
semiclassical treatment -- is general enough, so that, in principle, it can be applied to a large class of 
stochastic models. 

This paper is organized as follows. In Sec.~II, we briefly outline, as a ground for the subsequent 
discussions, the essential points of the RFDC model. In Sec.~III, we rephrase the RFDC model in the 
Martin-Siggia-Rose path-integral formalism and study it through the effective action method. 
Analytical expressions for the vgPDFs are then obtained. In Sec.~IV, we compare, using Monte-Carlo 
simulations, our vgPDFs with the ones derived from the numerical integration of the RFDC differential 
stochastic equations. Finally, in Sec.~V, we summarize our main findings and point out directions of 
further research.

\section{\leftline{The RFDC Lagrangian Stochastic Model}}

Our central object of interest is the time-dependent lagrangian velocity gradient tensor $\A(t)$, 
which has cartesian components $A_{ij} = \partial_i v_j$. Taking, as a starting point, the Navier-Stokes 
equations with external gaussian stochastic forcing, the exact lagrangian evolution equation for 
$\A(t)$ is
\be
\dot \A = V[ \A] + g\mathbb{F} \ , \ \label{lagrange_eq}
\ee
where  $V[\A]$ is a functional of $\A$, defined as
\be
V_{ij}[\A] = - (\A^2)_{ij} + \partial_i \partial_j \nabla^{-2} \tr (\A^2) + \nu \nabla^2 (\A)_{ij} \ , \ \label{V_exact}
\ee
and $\mathbb{F}$ is a zero-mean, second order gaussian random tensor, which satisfies to
\be
\langle F_{ij}(t) F_{kl}(t') \rangle = G_{ijkl}  \delta(t-t') \ , \
\ee
with
\be
G_{ijkl}=2 \delta_{ik} \delta_{jl} - \frac{1}{2} \delta_{il} \delta_{jk}- \frac{1}{2} \delta_{ij} \delta_{kl} \ . \
\ee
In  Eq.~(\ref{lagrange_eq}), $g$ is just an arbitrary coupling constant proportional to the external power per unit mass, 
which has an important role in our discussion, since it will be taken as an expansion parameter around the linearized model.

The second and third contributions to the right hand side of (\ref{V_exact}) are, respectively, the pressure Hessian 
(written as a non-local functional of the velocity gradient tensor) and the viscous dissipation term. As it stands,
 Eq.~(\ref{V_exact}) is of course not closed: exact solutions on a single lagrangian trajectory are clearly dependent 
on the bulk space-time profiles of the velocity gradient tensor. However, motivated by the fact that $\A(t)$ is typically 
short-time correlated, it is natural to conjecture that both the pressure Hessian and the viscous dissipation term are 
dominated by local contributions. This is the point of view taken in the RFDC model of Chevillard and Meneveau \cite{chevi_menev}, where these local contributions are related to the Kolmogorov and the large eddy time scales of
the flow, $\tau$ and $T$, respectively (the Reynolds number is, thus, $R_e \propto (T/\tau)^2$). It is then assumed 
that the lagrangian evolution of $\A(t)$ is associated, for small time scales, to the approximate Cauchy-Green tensor
\be
\C = \exp [ \tau \A] \exp [ \tau \AT ] \ , \ 
\ee
so that the functional $V[\A]$ in  Eq.~(\ref{lagrange_eq}) gets replaced by a local function of $\A$,
\be
V(\A) = - \A^2 +  \frac{\C^{-1} \tr( \A^2) }{\tr(\C^{-1})} - \frac{\tr(\C^{-1})}{3T} \A \ . \ \label{v(A)}
\ee
We end up, therefore, with a closed and much simpler time evolution equation for $\A$:
\be
\dot \A = V( \A) + g\mathbb{F} = - \A^2 +  \frac{\C^{-1} \tr( \A^2) }{\tr(\C^{-1})} - \frac{\tr(\C^{-1})}{3T} \A + g \mathbb{F} 
\ . \ \label{lagrange_eqb}
\ee
We refer the reader to Ref. \cite{chevi_menev} for a more detailed account on the conceptual and technical aspects 
of the RFDC model.

It is convenient to set $T=1$ (without loss of generality) and perform an expansion of 
$V(\A)$ up to some arbitrary power of $\tau$ in (\ref{v(A)}). A previous extensive numerical study
shows that even the first order expansion is enough to grasp the physical content of the model \cite{afonso_menev}. 
We work, throughout the paper, with second order expansions of $V(\A)$, which we write as
\be
V(\A) = \sum_{p=1}^4 V_p(\A)  \ , \ \label{Vsum}
\ee
where
\bea
&&V_1(\A) = - \A \ ,\ \\
&&V_2(\A)= - \A^2 +\frac{\I}{3} \tr(\A^2) +\frac{2 \tau }{3} \tr(\A) \A \ , \ \\
&&V_3(\A) = -\frac{\tau}{3} \left (\A +\AT - \frac{2\I}{3} \tr(\A) \right ) \tr (\A^2) - \frac{\tau^2}{3} \tr (\AT \A) \A -\frac{\tau^2}{3} \tr (\A^2) \A \ , \ \\
&&V_4(\A) = -\frac{\I}{9} \tau^2 \tr (\AT \A) \tr(\A^2) - \frac{\I}{9} \tau^2 [\tr(\A^2)]^2 + \frac{\tau^2}{3} \AT \A \tr(\A^2) + \nonumber \\
&&+  \frac{\tau^2}{6} ( \A^2 + \ATT) \tr(\A^2) \ . \ 
\eea
It is important to note that $V_p(\A)$ comprises all the contributions of $O(\A^p)$ to $V(\A)$. The RFDC model yields a promising stage for further improvements, insofar its vgPDFs as well as its geometrical statistical properties related to the coupling between the vorticity and the rate-of-strain tensor share several qualitative features in common with the ones observed in experiments and direct numerical simulations of turbulence \cite{chevi_menev2,chevi_etal}.

\section{Path-Integral Formulation of the RFDC Model}

Assume that at time $t=0$ the velocity gradient tensor is $\A(0) \equiv \A_0$. We may write, within the framework of the Martin-Siggia-Rose (MSR) functional formalism \cite{msr,janssen,dominicis,cardy}, the following path-integral expression for the conditional probability density function of finding, at time $t=\beta$, the velocity gradient tensor $\A(\beta) \equiv \A_1$,
\be
\rho(\A_1 | \A_0, \beta) \equiv {\cal{N}} \int_\Sigma  D[\hat \A] D[\A] \exp \left \{ -\int_0^\beta dt \left [ i \tr [ \hAT L(\A) ] + \frac{g^2}{2} G_{ijkl} \hat A_{ij} \hat A_{kl}  \right ] \right \} \ , \ \label{vgpdf_cond}
\ee
where 
\be
\Sigma = \{ \A(0)=\A_0 \ , \ \A(\beta) = \A_1 \} \label{sigma}
\ee
specifies the set of boundary conditions, $\cal{N}$ is a normalization factor, and 
\be
L(\A) \equiv \dot \A - V(\A) \ . \
\ee
In the above expression, $\hat \A = \hat \A(t)$ is just an auxiliary tensor field (a time-dependent $3 \times 3$ matrix) with no direct physical meaning. The conditional PDF given in  Eq.~(\ref{vgpdf_cond}) is nothing but a formal solution, written with path-integral dressing, of the Fokker-Planck equation that can be derived from the stochastic differential  Eq.~(\ref{lagrange_eqb}). 

\vspace{0.3cm}

{\leftline{\it{General Strategy for the derivation of vgPDFs}}}
\vspace{0.3cm}

Taking $\beta \rightarrow \infty$ in the conditional PDF (\ref{vgpdf_cond}), we obtain the stationary vgPDF evaluated at $\A_1$, which is expected to be independent from the initial condition $\A_0$. Also, as it is clear from the original RFDC equations, in the limit of small $g$, nonlinear perturbations become negligible and all we get are gaussian distributions for the vgPDFs. We are, thus, interested to investigate how the vgPDFs evolve as the noise strength $g$ gets progressively larger and intermittency effects cannot be neglected anymore. Straightforward semiclassical evaluations can be implemented, in principle, along two alternative procedures \cite{amit,zinn-justin}: (i) in the WKB approach, quadratic fluctuations around the classical Euler-Lagrange equations of motion are tentatively integrated; (ii) in the Effective Action method, the path-integral can be evaluated up to some order in perturbation theory and Euler-Lagrange equations are then subsequently studied.

In both methods (i) and (ii), it is necessary to deal with variational Euler-Lagrange equations which take into account the boundary conditions in $\Sigma$, as defined in (\ref{sigma}). However, it is important to note that while in (i) the integration over quadratic fluctuations boils down to the computation of a functional determinant, the perturbative expansion in (ii) is generally associated to the evaluation of one-loop Feynman diagrams. As it is well-known from standard field-theoretical arguments \cite{zinn-justin}, the computation of the functional determinants in the WKB approach encodes, in general, a complet set of one-loop one-particle irreducible (1PI) Feynman diagrams, which turn out to be, in our case, of unfortunate cumbersome implementation. Within the Effective Action method, on the other hand, we can select the one-loop diagrams which we assume are the most relevant and check, {\it{a posteriori}}, how good will this selection perform. This is the pragmatic point of view that we follow throughout this work.

To start, we take profit of the independence of the conditional vgPDF (\ref{vgpdf_cond}) upon the initial condition $\A_0$, for $\beta \rightarrow \infty$, and impose the particular periodic boundary conditions 
\be
\A(0) = \A(\beta) = \bar \A \ . \
\ee
As it will be clear a bit later, the choice of periodic boundary conditions for $\A(t)$ is very convenient, once it leads to great simplifications in the structure of the effective action. We may state, therefore, that up to an unimportant renormalization factor (which from now on is suppressed, for convenience, from all the PDF expressions),
\be
\rho(\bar \A) \equiv \lim_{\beta \rightarrow \infty} \rho(\bar \A|\bar \A, \beta) = \exp \{- S_c[\hat \A^c , \A^c] \} \label{vgPDF}
\ee
is the probability density function of having $\A(t) = \bar \A$ at an arbitrary time instant $t$ in the asymptotic
stationary fluctuation regime. In (\ref{vgPDF}), $S_c[\hat \A^c , \A^c]$ is the ``Effective Martin-Siggia-Rose 
Action", which satisfies to
\bea
&&\left. \frac{\delta S_c[\hat \A^c , \A] }{\delta A_{ij}} \right |_{\A = \A^c} = 0 \label{speq1} \ , \ \\
&&\left. \frac{\delta S_c[\hat \A , \A^c] }{\delta \hat A_{ij}} \right |_{\hat \A = \hat \A^c} = 0  \label{speq2} \ , \
\eea
subject to the boundary conditions $\A^c(0)=\A^c(\beta)=\bar \A$. The index $c$ used in Eqs.~(\ref{vgPDF}-\ref{speq2}) stands for 
``classic", following the spread field-theoretical jargon.

If we are able to handle the Euler-Lagrange Eqs.~(\ref{speq1}) and (\ref{speq2}) to write $\hat \A$ in terms of $\A$, 
then we may define
\be
S_c[\A] \equiv S_c[\hat \A(\A),\A] \ . \ 
\ee
We have, therefore,
\be
\left. \frac{\delta S_c[\A]}{\delta A_{ij}} \right |_{\A^c} = \left. \frac{\delta S_c[\hat \A^c,\A]}{\delta A_{ij}} \right |_{\A = \A^c}
+\left. \frac{\delta S_c[\hat \A, \A^c]}{\delta \hat A_{kl}} \right |_{\hat \A = \hat \A^c} \times \left. \frac{\delta \hat A_{kl}}{\delta A_{ij}} \right |_{\A = \A^c} = 0 \ . \  \label{speq}
\ee
In other words, in order to find the vgPDF (\ref{vgPDF}), it is necessary to solve just for one extrema $\A^c(t)$ of effective action $S_c[\A]$, as obtained from  Eq.~(\ref{speq}), instead of considering, in an explicit way, the solutions of both Eqs.~(\ref{speq1}) and (\ref{speq2}).

The effective MSR action can be systematically evaluated to arbitrary degrees of precision. As a working hypothesis, we focus our attention only on the effects of noise renormalization, which amounts to say that the kernel $G_{ijkl} \delta (t-t')$, implicit in (\ref{vgpdf_cond}), gets substituted by a corrected one, $G^{ren}_{ijkl}(t-t')$, so that
\be
S_c[\hat \A , \A] = i \int_0^\beta dt  \tr [ \hAT L(\A) ] + \frac{g^2}{2} \int_0^\beta dt \int_0^\beta dt' G^{ren}_{ijkl}(t-t') \hat A_{ij}(t) \hat A_{kl} (t') \label{effective_action} \ . \
\ee
The saddle-point equations (\ref{speq1}) and (\ref{speq2}) give, respectively,
\bea
&&i \dot {\hat A}_{ij} + \hat A_{kl} \frac{\partial [V(\A)]_{kl}}{\partial A_{ij}} = 0 \ , \ \label{speq_hat_a}\\
&&iL_{ij}(\A) + g^2 \int_0^\beta dt' G^{ren}_{ijkl}(t-t') \hat A_{kl}(t') = 0 \label{speq_a} \ . \
\eea
It turns out that $G^{ren}_{iikl}(t-t')= G^{ren}_{ijkk}(t-t') = 0$ to all orders in perturbation theory, a result that is related to the fact that 
$\tr(\A)=0$ for the solutions of the RFDC equations. This leads, from  Eq.~(\ref{speq_a}), to $\tr[L(\A)] =0$. 

As it is usually done in effective action studies \cite{amit, zinn-justin}, we work with low-frequency 
renormalization, once the saddle-point solutions are assumed to couple in a weak way to the fast degrees of
freedom. We just mean here the replacement of the operator kernel $G^{ren}_{ijkl}(t-t')$ by 
$\tilde G^{ren}_{ijkl} \delta(t-t')$, where
\be
\tilde G^{ren}_{ijkl} \equiv \int_{-\infty}^{\infty} dt G^{ren}_{ijkl}(t) \ . \
\ee
Eq.~(\ref{speq_a}) is transformed, in this way, into the local equation
\be
iL_{ij}(\A) + g^2 \tilde G^{ren}_{ijkl} \hat A_{kl} = 0  \label{speq_local} \ . \
\ee
Taking into account now that $G^{ren}_{iikl} = G^{ren}_{ijkk} = 0$, we define,
\be
\tilde G^{ren}_{ijkl} \equiv D_{ijkl} - \frac{1}{3} (x+y) \delta_{ij} \delta_{kl} \ , \ \label{xy1} 
\ee
where $x$ and $y$ are computable parameters, and
\be
D_{ijkl} \equiv x \delta_{ik} \delta_{jl} +y \delta_{il} \delta_{jk} \ , \ \label{xy2}
\ee
then  Eq.~(\ref{speq_local}) can be rewritten as
\be
g^2 D_{ijkl} \hat A_{kl} = -iL_{ij}(\A) + \frac{g^2}{3}(x+y) \tr[\hat \A] \delta_{ij} \ , \
\ee
which leads to 
\be
\hat A_{ij} = -\frac{i}{g^2} D^{-1}_{ijkl} L_{kl}(\A)+ \frac{1}{3}\tr[ \hat \A]\delta_{ij} \label{hat_a} \ , \
\ee
a traceless tensor. Substituting Eq.~(\ref{hat_a}) in the effective action (\ref{effective_action}), 
we get, in the low-frequency regime,
\be
S_c[\A] = \frac{1}{2g^2} 
\int_0^\beta dt \left [ D^{-1}_{ijkl}L_{ij}(\A) L_{kl}(\A)  \right ] \label{speq_ren} \ . \
\ee
Note that it is not difficult at all to obtain $D^{-1}_{ijkl}$. Defining
\be
D^{-1}_{ijkl} \equiv a \delta_{ik}\delta_{jl} + b  \delta_{il}\delta_{jk} \ , \ \label{invD}
\ee
the tensorial equation $D^{-1}_{ijpq}D_{pqkl} = \delta_{ik}\delta_{jl}$ implies, after some simple algebra, that
\be
\left\{
\begin{array}{l}
ax + by = 1  \\
ay + bx = 0
\end{array}
\right.  \ , \ 
\ee
and, consequently,
\be
a = - \frac{x}{y^2-x^2} \ , \ b = \frac{y}{y^2-x^2} \ . \ \label{ab}
\ee
Our general strategy, is, thus, very clear: we have to determine, firstly, the values of $x$ and $y$, as they are introduced in the definitions (\ref{xy1}) and (\ref{xy2}) for the renormalized noise kernel. In second place, we study the saddle point solutions of the effective action (\ref{speq_ren}). Once we have these ingredients at hand, we can immediately write down an expression for the vgPDF in the RFDC model, viz,
\be
\rho(\bar \A) = \exp \{- S_c[\A^c] \} \ , \  \label{vgPDF_b}
\ee
where $S_c[ \A^c]$ is given by  Eq.~(\ref{speq_ren}), with $\A(t)$ substituted by $\A^c(t)$.
\vspace{0.3cm}

{\leftline{\it{Noise Renormalization}}}
\vspace{0.3cm}

The functional Taylor expansion of the effective action is obtained from the evaluation of operator kernels, 
which are identified to $N$-point 1PI Feynman diagrams \cite{amit,zinn-justin}. They have, as a general rule, a prefactor
that is proportional to the perturbative coupling constant ($g^2$ in our case) raised to a power which is directly related 
to the number of loops that each 1PI diagram contains.

The perturbative expansion is achieved very straightforwardly, in the path integral formulation, 
through the power series expansion of the exponentiated non-quadratic terms of the MSR action. 
The quadratic terms in the MSR action define the ``free theory", where arbitrary expectation values, 
represented by the notation $\langle (...) \rangle_0$, can be exactly evaluated. In our particular 
problem, the free one-particle propagator is defined as
\be
\langle A_{ij}(t) \hat A_{kl}(t') \rangle_0 = i \delta_{ik} \delta_{jl} G_0(t-t')
\ee
where
\be
G_0(t-t') \equiv \Theta(t-t') \exp (t'-t)
\ee
is the Green's function of the differential operator
\be
D_t \equiv 1 + \partial_t \ . \
\ee 
The renormalized noise kernel $G^{ren}_{ijkl}(t-t')$ is obtained from the relation
\be
D_t D_{t'}\langle A_{ij}(t) A_{kl}(t') \rangle_{1PI} \equiv  g^2 G^{ren}_{ijkl}(t-t') \ , \ \label{DD}
\ee
where $\langle (...) \rangle_{1PI}$ stands for the use of only 1PI contributions in the perturbative expansion
of the expectation value under consideration. We have, up to $O(g^4)$, 
\be
D_t D_{t'}\langle A_{ij}(t) A_{kl}(t') \rangle_{1PI} = g^2 [G_{ijkl} \delta(t-t') + g^2 C_{ijkl}(t-t')] \ , \ \label{noise_1PI}
\ee
where
\be
C_{ijkl}(t-t') = \frac{1}{8}G_{abcd}G_{efgh} \int d \xi d\xi' \langle
 [V_2(\A(t))]_{ij} [V_2(\A(t'))]_{kl} \hat A_{ab}(\xi) \hat A_{cd}(\xi)
 \hat A_{ef}(\xi') \hat A_{gh}(\xi') \rangle_0 \ , \
\ee
which means that $G^{ren}_{ijkl}(t-t') = G_{ijkl}\delta(t-t') + C_{ijkl}(t-t')$. 
The two contributions in the right-hand-side of (\ref{noise_1PI}) are represented by the two diagrams
depicted in Fig.~1 \cite{barabasi-stanley}. The dashed lines in these diagrams are the external propagator 
lines which are removed in the 1PI contributions due to the action of the differential operators 
$D_t$ and $D_{t'}$ in Eq.~(\ref{DD}).

\begin{center}
\begin{figure}
\vspace{-7.0cm}
\hspace{-2.0cm}
\includegraphics[width=12.8cm, height=18.4cm]{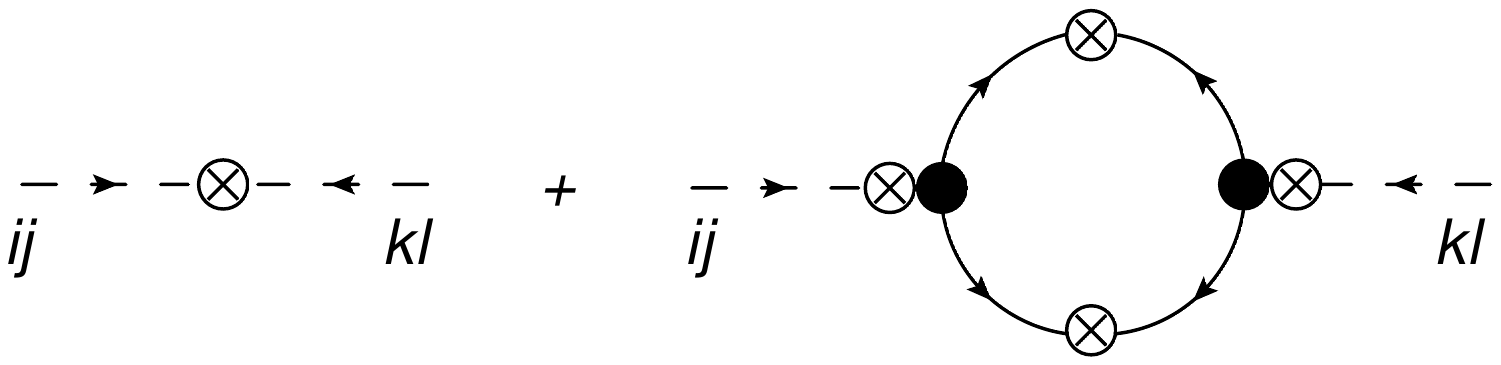}
\label{}
\vspace{-7.5cm}
\caption{The 1PI Feynman diagram contributions to the noise vertex renormalization
of the effective Martin-Siggia-Rose action, up to O($g^4$). The free one-particle 
propagators are represented by directed lines. The bare noise vertices are depicted as
isolated crossed circles linked to two convergent lines. The bare three-point vertices 
are associated to the contributions provided by $V_2(\A)$ in the RFDC stochastic 
time evolution equation (\ref{lagrange_eqb}).}
\end{figure}
\end{center}

Recalling the notation introduced in (\ref{xy1}) and (\ref{xy2}), we find, after straightforward
computations,
\bea
&&x = 2+ \frac{3}{2} g^2 \ , \ \\
&&y = -\frac{1}{2} -\frac{1}{16} g^2 \ . \
\eea
Taking into account, now, (\ref{invD}) and (\ref{ab}), the effective action (\ref{speq_ren}) can be written, 
therefore, as
\be
S_c[\A] = \frac{1}{2g^2} 
\int_0^\beta dt \left \{ a \tr \left [ L^{\hbox{\scriptsize{T}}}(\A) L(\A) \right ]  +  b \tr \left [L^2(\A) \right ]  
\right \} \label{speq_ren2} \ . \
\ee
\vspace{0.3cm}

{\leftline{\it{Saddle-Point Solutions}}}
\vspace{0.3cm}

It is a hard -- if not an actually impossible -- task to obtain the exact saddle-point solutions of the Euler-Lagrange equations (\ref{speq}) derived from the effective action (\ref{speq_ren2}). However, we can in principle retain, in the small $g$ regime, only the quadratic terms in the effective action and use the saddle-point solutions obtained in this way as a first approximation to the exact solutions for larger values of $g$. We have, in 
the quadratic approximation,
\be
S_c[\A] \equiv  \frac{a}{2g^2} \int_0^\beta dt  {\hbox{Tr}}  \left [\dot \AT \dot \A + \AT \A \right ] 
+ \frac{b}{2g^2} \int_0^\beta dt  {\hbox{Tr}} \left [ \dot \A^2 + \A^2 \right ] \ . \
\ee
The saddle-point  Eq.~(\ref{speq}) yields, in this case,
\be
\ddot \A - \A = 0 \ . \ \label{Aeq}
\ee
The solution of (\ref{Aeq}) that satisfies the boundary conditions (\ref{sigma}) is
given by
\be
\A(t) = \bar \A f_\beta(t) \ , \ \label{Af}
\ee
where
\be
f_\beta(t) = 2 \frac{\sinh( \frac{\beta}{2} )}{\sinh(\beta)} \cosh(t - \frac{\beta}{2}) \ . \
\ee
Substituting (\ref{Af}) in (\ref{Vsum}), we obtain
\be
V(\A(t)) = \sum_{p=1}^4 V_p(\bar \A) [f_\beta(t)]^p  \ . \
\ee
The effective MSR action becomes, in the limit where $\beta \rightarrow \infty$,
\be
S_c[\A^c] = S(\bar \A)= S_1(\bar \A) +S_2(\bar \A) \ , \
\ee
with
\be
S_1(\bar \A) = \frac{a}{2 g^2} \tr \left [ I_1 \barAT \bar \A +
\sum_{p=1}^4 \sum_{q=1}^4  I_{p+q} V_p(\barAT)V_q(\bar \A) \right ] \ , \
\ee
and
\be
S_2(\bar \A) = \frac{b}{2 g^2} \tr \left [ I_1 {\bar \A}^2 +
\sum_{p=1}^4 \sum_{q=1}^4  I_{p+q} V_p(\bar \A)V_q(\bar \A) \right ] \ , \
\ee
where the above $I$-coefficients are defined as
\be
I_1 = \lim_{\beta\rightarrow \infty} \int_0^\beta dt [ \dot f_\beta(t)]^2 \ , \ I_{p+q} = \lim_{\beta\rightarrow \infty} \int_0^\beta dt [f_\beta(t)]^{p+q} \ . \
\ee
Their numerical values are listed below
\be
I_1=I_2=1 \ , \ I_3 = 2/3 \ , \ I_4 = 1/2 \ , \ I_5 = 2/5 \ , \ I_6 = 1/3 \ , \ I_7 = 2/7 \ , \ I_8 = 1/4 \ . \
\ee
We emphasize, at this point, that the number of terms that contribute to $S(\bar \A)$ would be unnecessarily larger
had we not used periodic boundary conditions for $\A^c(t)$. The reason is that due to the periodic boundary
conditions, the several time integrations of tensorial products involving only one time derivative of the 
velocity gradient tensor can be removed from the effective action evaluated at its saddle-point configurations.

\section{Analytical versus Empirical ${\hbox{\small{vg}}}$PDF${\hbox{\small{s}}}$}

Plots of the analytical vgPDFs $\rho(\bar \A) \propto \exp [ -S(\bar \A) ]$ can be compared to the empirical PDFs obtained
through the direct numerical solutions of the stochastic differential  Eq.~(\ref{lagrange_eqb}). We have produced, using the analytical vgPDFs,
large Monte Carlo ensembles of velocity gradients. The numerical solution of  Eq.~(\ref{lagrange_eqb}), on the other hand 
is carried out within a second order predictor-corrector method \cite{kloeden-platen}, with time step $\epsilon = 0.01$. We have considered, 
in all our numerical tests, $\tau = 0.1$, a reference time-scale usually taken in studies of the RFDC model.

In our Monte Carlo procedure, the velocity gradient $\A$ is additively perturbed by random traceless $3 \times 3$ matrices at each iteration step. 
The stochastic increments can be always written as a linear superposition of matrices of the overcomplete set $\{ \B_1, \B_2,...,\B_9 \}$, where
\bea
&& \B_1 = 
\left[ \begin{array}{ccc}
0 & 0 & 0 \\
0 & 0 & 1 \\
0 & -1 & 0
\end{array} \right]
\ , \
\B_4 =
\left[ \begin{array}{ccc}
0 & 1 & 0\\
1 & 0 & 0\\
0 & 0 & 0
\end{array} \right]
\ , \
\B_7 =
\left[ \begin{array}{ccc}
1 & 0 & 0\\
0 & -\frac{1}{2}  & 0\\
0 & 0 &  -\frac{1}{2}
\end{array} \right] \ , \ \nonumber  \\
&&\B_2 = 
\left[ \begin{array}{ccc}
0 & 0 & 1 \\
0 & 0 & 0 \\
-1 & 0 & 0
\end{array} \right]
\ , \
\B_5 = 
\left[ \begin{array}{ccc}
0 & 0 & 0\\
0 & 0 & 1\\
0 & 1 & 0
\end{array} \right]
\ , \
\B_8 =
\left[ \begin{array}{ccc}
-\frac{1}{2} & 0 & 0\\
0 & 1 & 0\\
0 & 0 &  -\frac{1}{2}  
\end{array} \right] \ , \ \nonumber \\
&&
\B_3 = 
\left[ \begin{array}{ccc}
0 & 1 & 0 \\
-1 & 0 & 0 \\
0 & 0 & 0
\end{array} \right]
\ , \
\B_6 =
\left[ \begin{array}{ccc}
0 & 0 & 1 \\
0 & 0 & 0 \\
1 & 0 & 0
\end{array} \right]
\ , \
\B_9 =
\left[ \begin{array}{ccc}
-\frac{1}{2} & 0 & 0\\
0 & -\frac{1}{2} & 0\\
0 & 0 &  1  
\end{array} \right] \ . \
\eea
Observe that the above matrices are special generators of three-dimensional rotations ($\B_1,\B_2,\B_3$), 
reflections ($\B_4,\B_5,\B_6$) and shearing transformations ($\B_7,\B_8,\B_9$). In more precise terms, 
the ensemble of velocity gradients is produced from successive stochastic perturbations of $\A$ given as
\be
\A \rightarrow \A' = \A + s \B_p \ . \ \label{iter}
\ee
Let $\Delta S(s) \equiv S(\A') - S(\A)$ and $\chi$ be a gaussian random variable which is sorted at each Monte Carlo step, with zero mean and some standard deviation $\sigma$, to be defined below. In order to set $s$ in (\ref{iter}), the Metropolis algorithm \cite{binder-landau} is then applied as follows:

(i) If $\Delta S(\chi) < 0$, take $s = \chi$, otherwise define $p = \exp[- \Delta S(\chi)]$ and go to step (ii).

(ii) Take $s = \chi$ with probability $p$ and $s=0$ with probability $1-p$.

\noindent As it is usually done in analogous Monte Carlo simulation contexts \cite{itzykson-drouffe}, the numerical value of the standard deviation parameter $\sigma$ is adjusted so that the case $s=0$ is verified in about $50\%$ of the Monte Carlo steps.

\begin{center}
\begin{figure}
\vspace{-0.0cm}

\includegraphics[width=7.8cm, height=5.39cm]{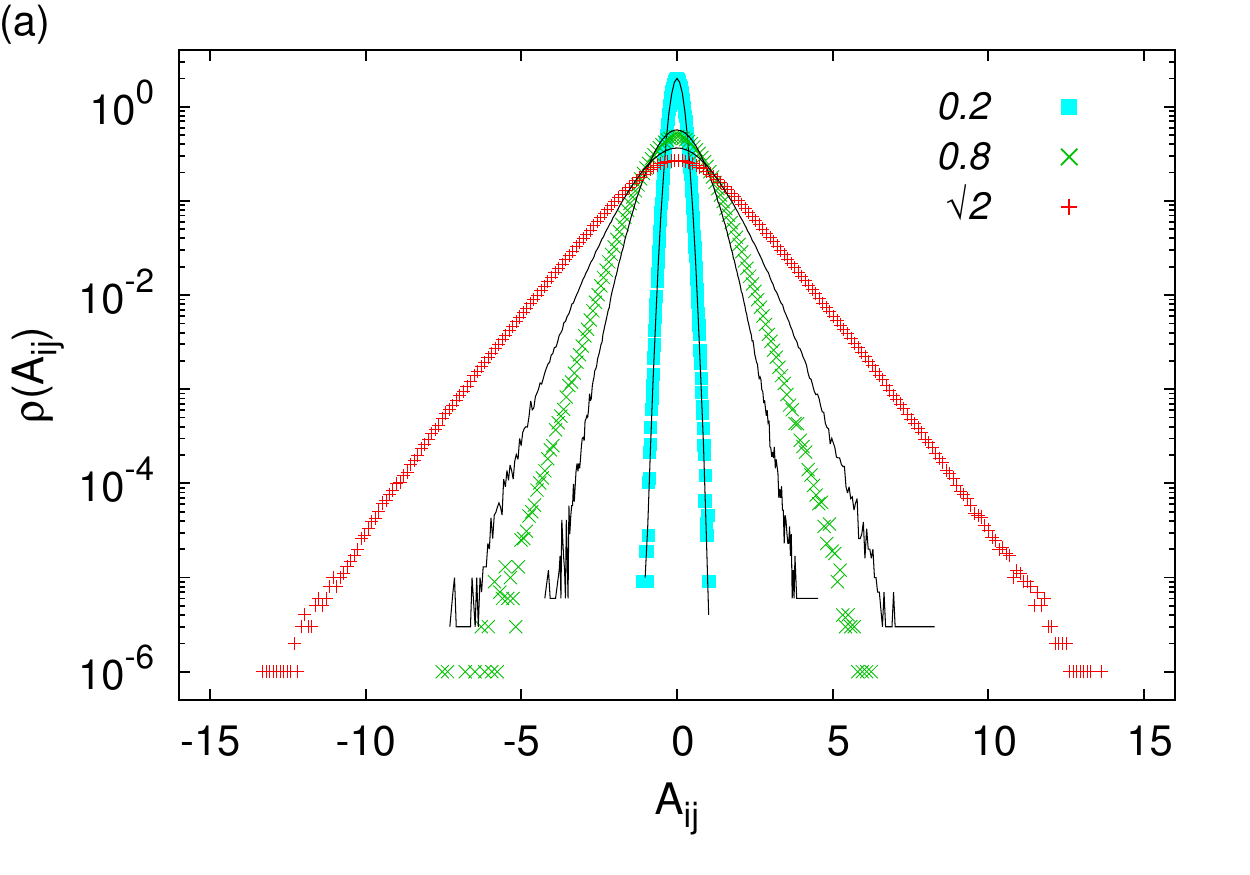}
\includegraphics[width=7.8cm, height=5.39cm]{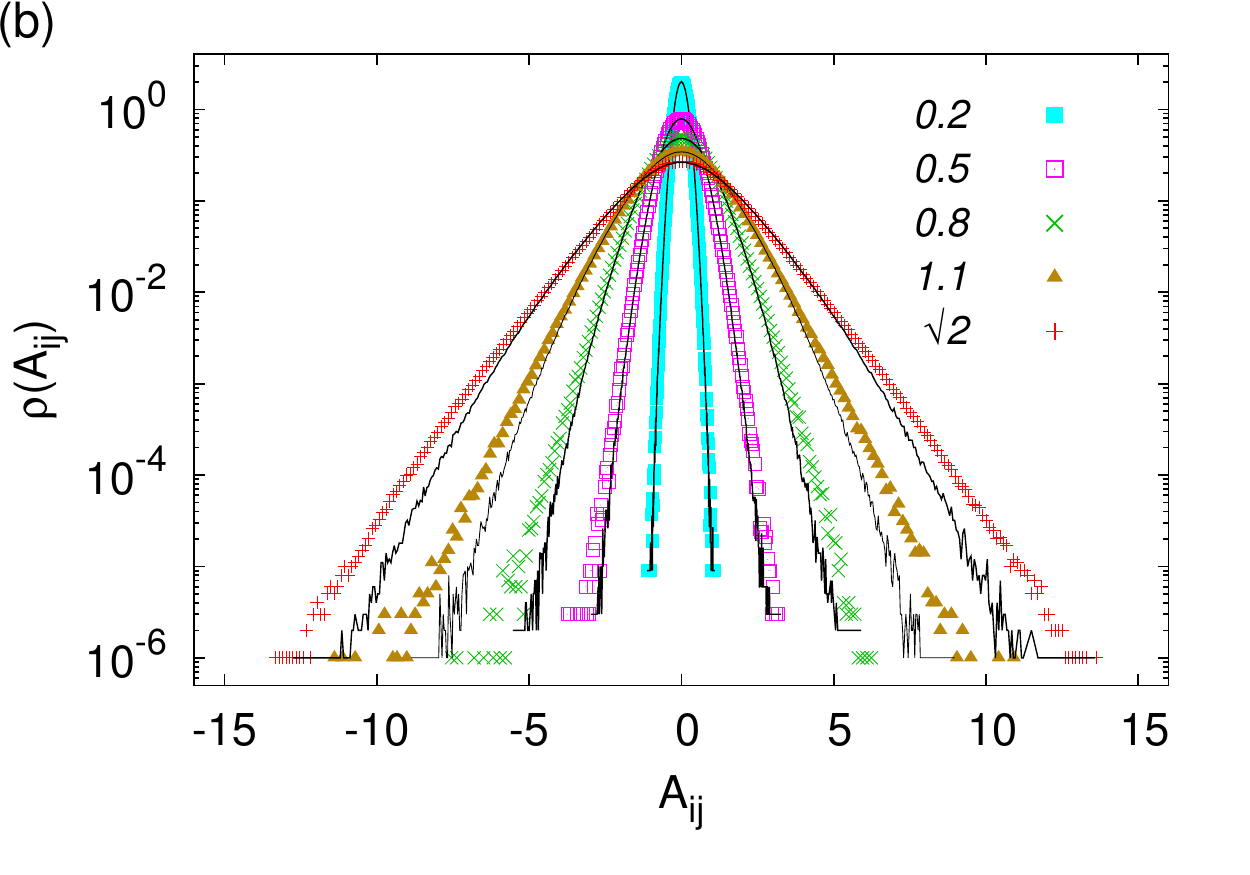}
\includegraphics[width=7.8cm, height=5.39cm]{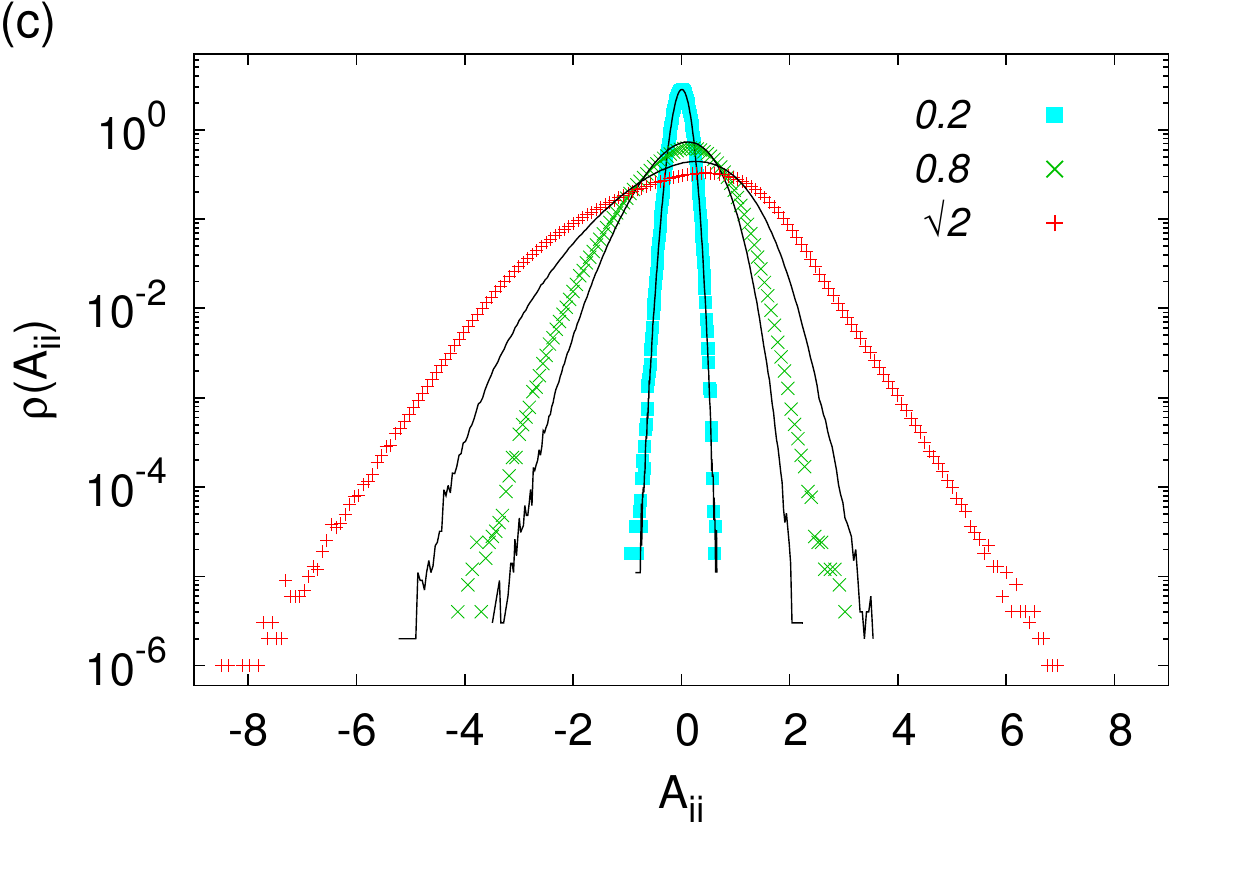}
\includegraphics[width=7.8cm, height=5.39cm]{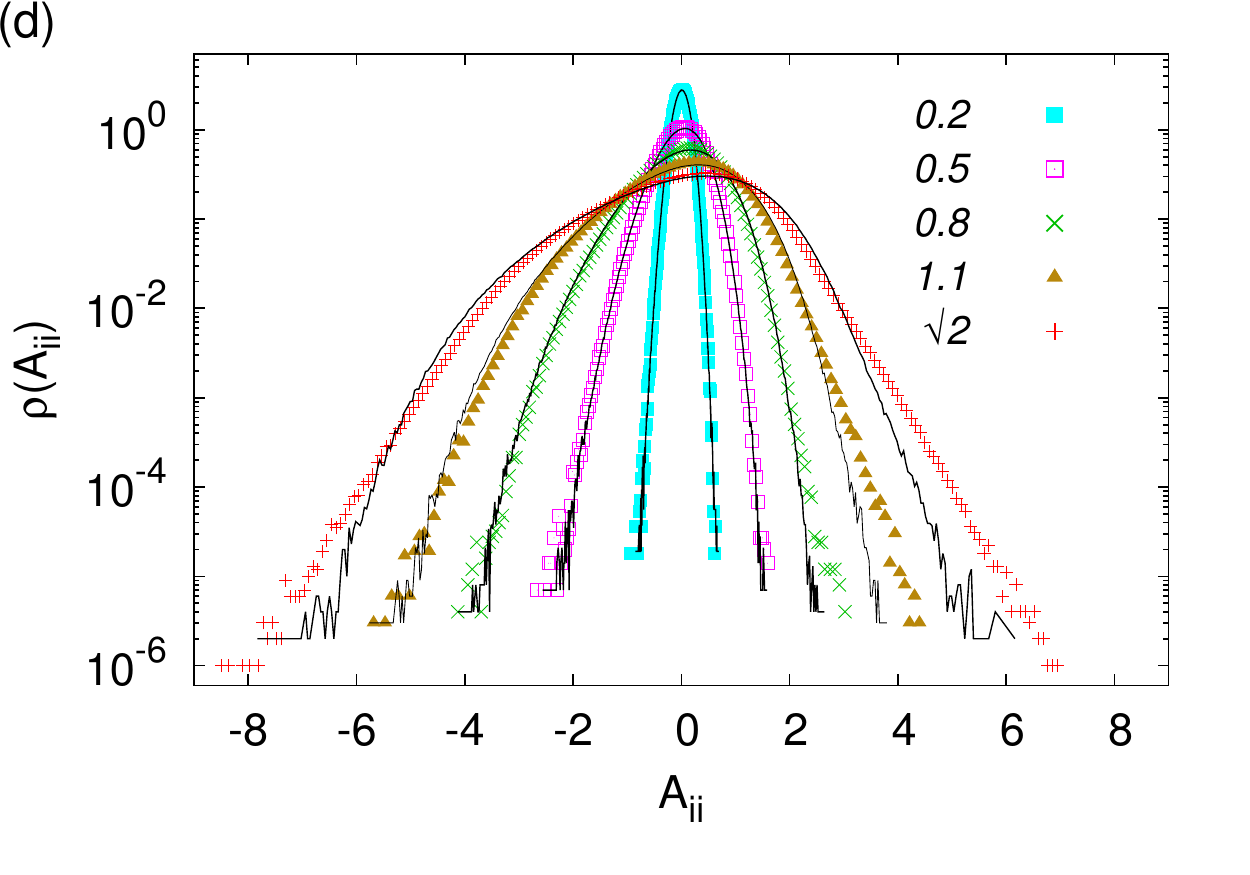}

\label{}
\vspace{-0.0cm}

\caption{Comparative semi-log plots of vgPDFs. The solid lines represent the 
analytical vgPDFs evaluated for $\tau=0.1$ and some values of the bare noise 
strength $g$ with and without noise renormalization. The vgPDFs depicted with 
symbols refer to the ones obtained from the direct numerical integration of 
the RFDC stochastic equations. The vgPDFs for the non-diagonal components of 
the velocity gradient tensor are given in figures (a) (nonrenormalized noise 
for g = 0.2, 0.8, and $\sqrt{2}$) and (b) (renormalized noise for g = 0.2, 
0.5, 0.8, 1.1 and $\sqrt{2}$). The analogous results for the diagonal 
components are given in figures (c) and (d).}
\end{figure}
\end{center}

\begin{center}
\begin{figure}
\vspace{-0.0cm}

\includegraphics[width=7.8cm, height=5.39cm]{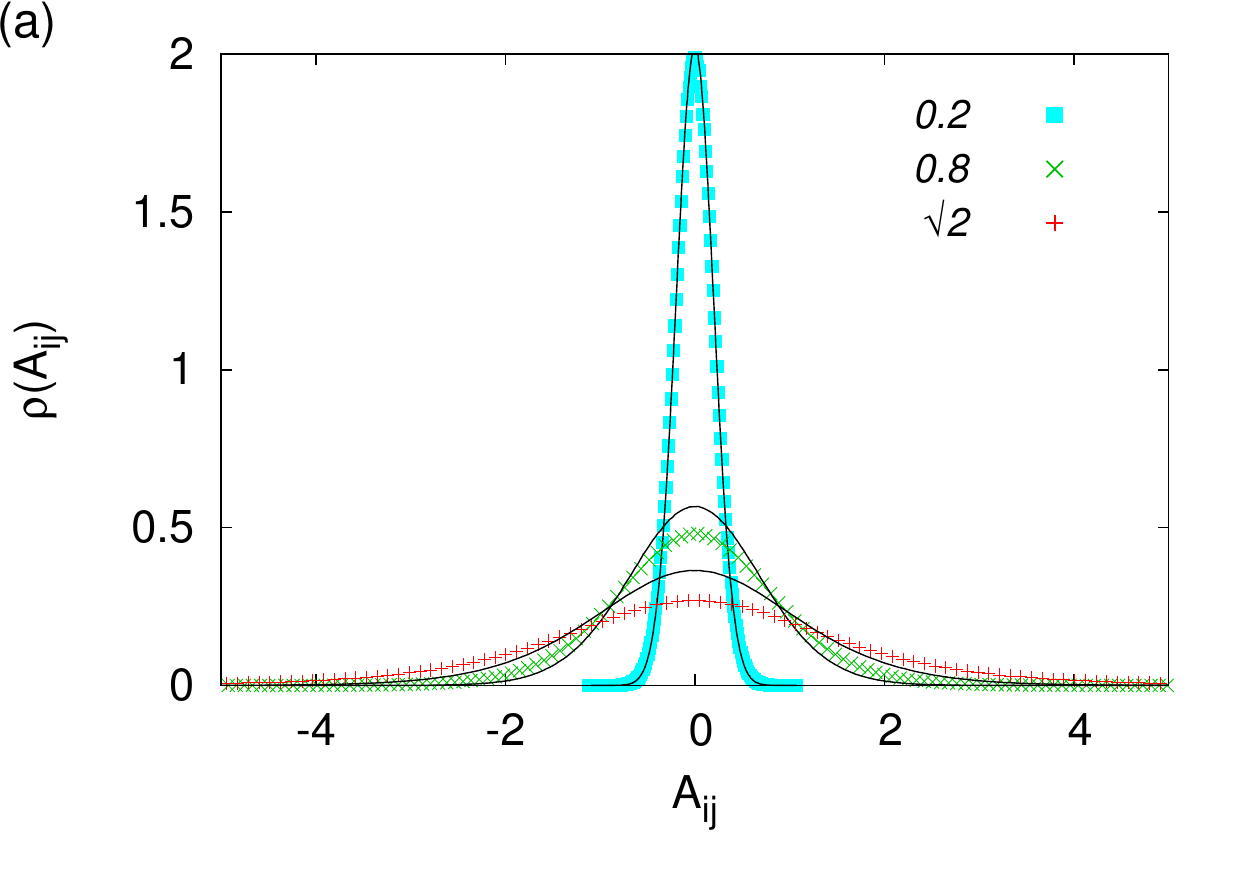}
\includegraphics[width=7.8cm, height=5.39cm]{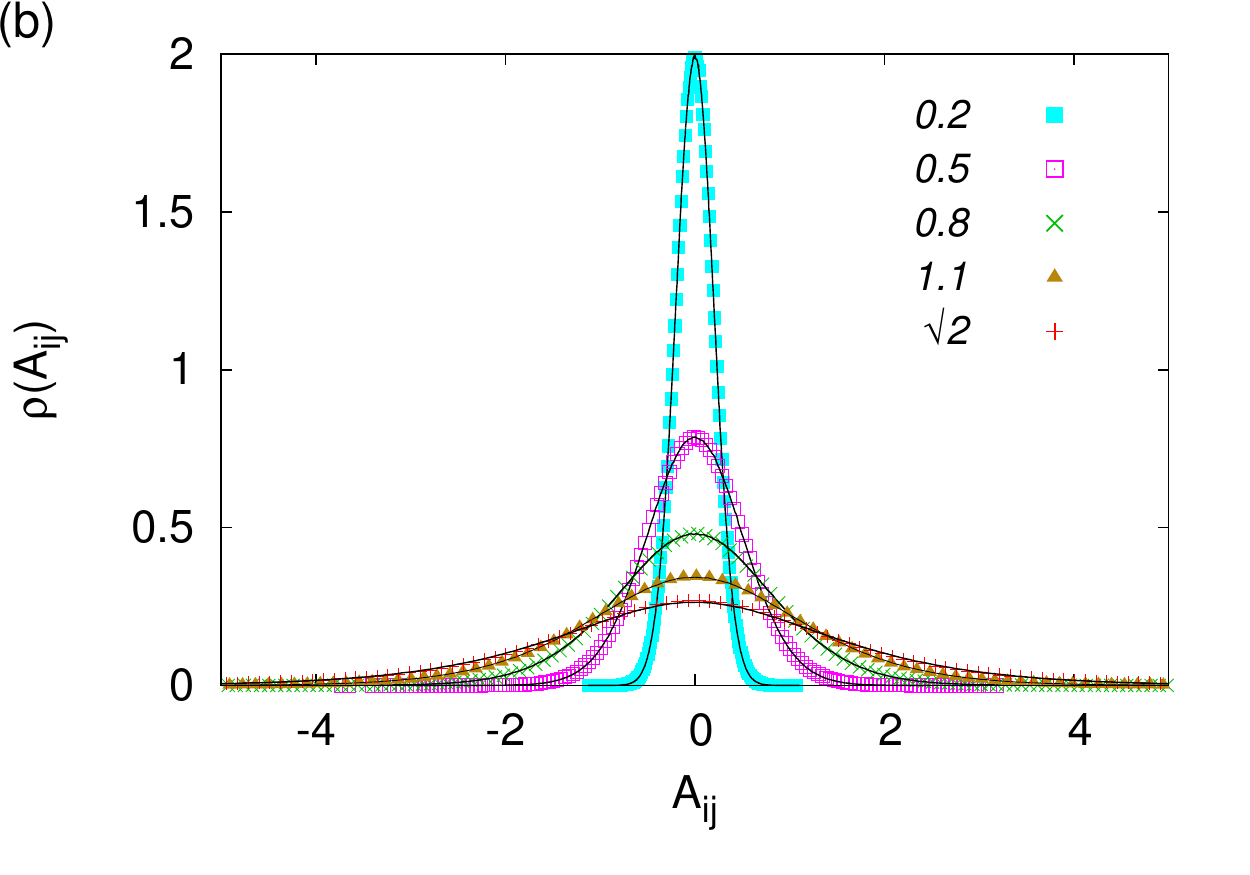}
\includegraphics[width=7.8cm, height=5.39cm]{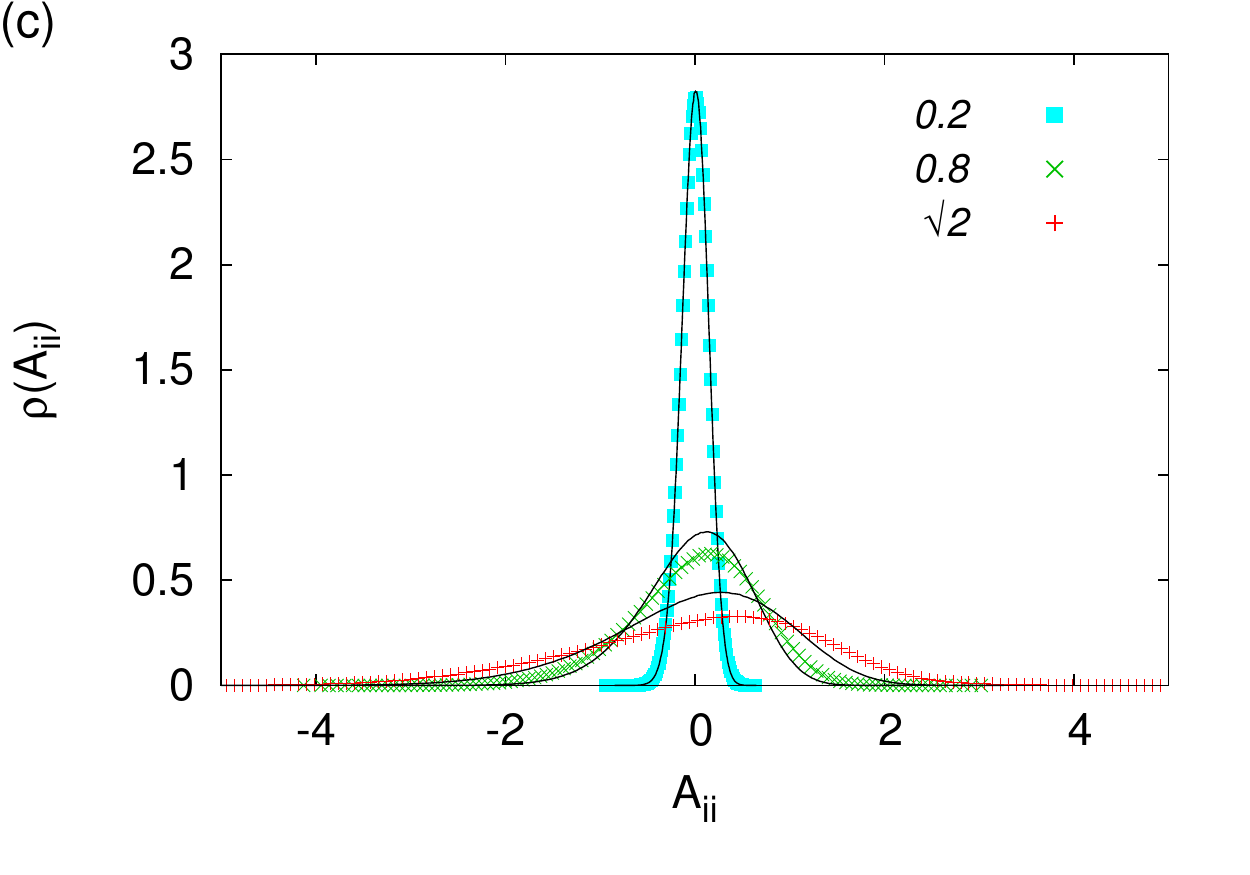}
\includegraphics[width=7.8cm, height=5.39cm]{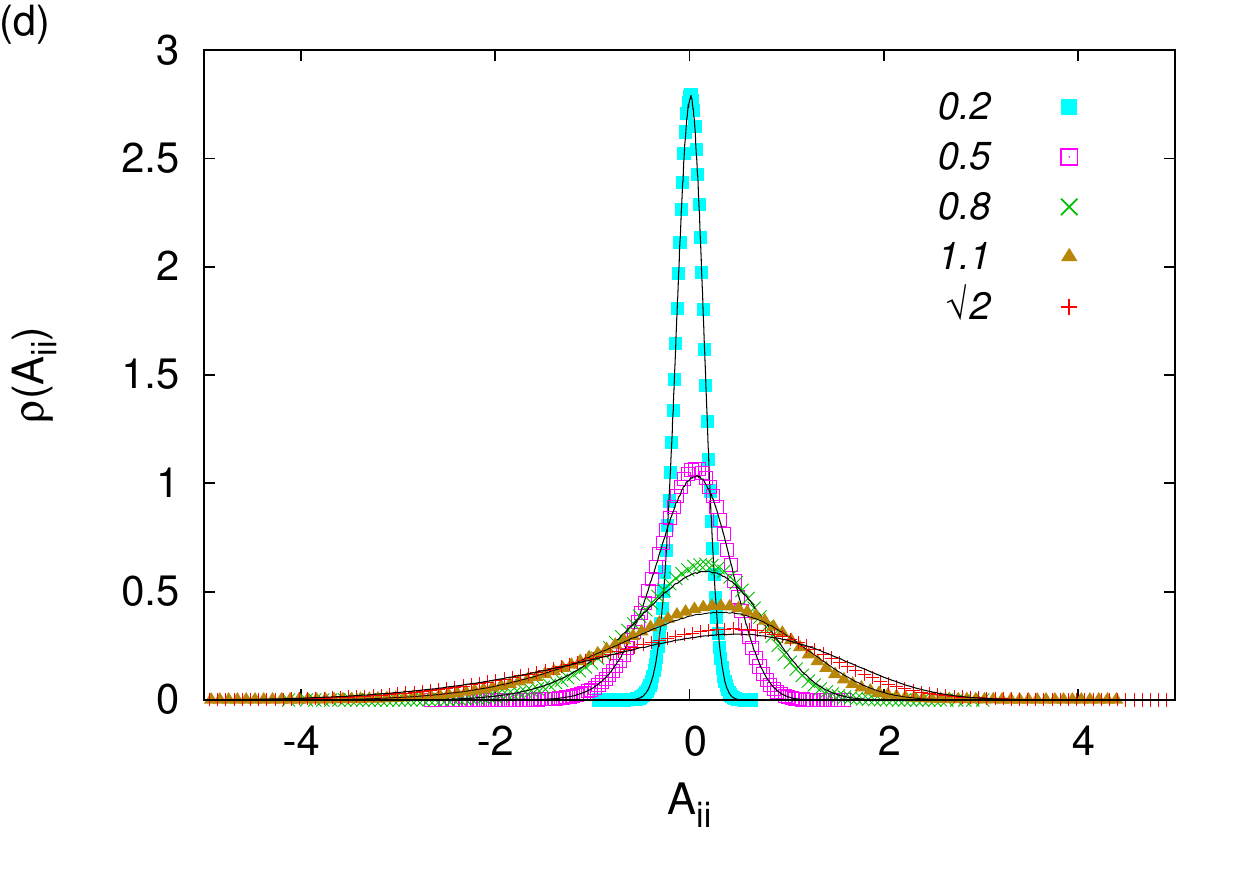}
\label{}
\vspace{0.0cm}

\caption{Comparative linear plots of vgPDFs, described
in the same way as in Fig.~2. The fittings in (b) and (d) 
are very reasonable within about two standard deviations
around the peak values of the vgPDFs.}
\end{figure}
\end{center}

\begin{center}
\begin{figure}
\vspace{-0.0cm}

\includegraphics[width=7.8cm, height=5.39cm]{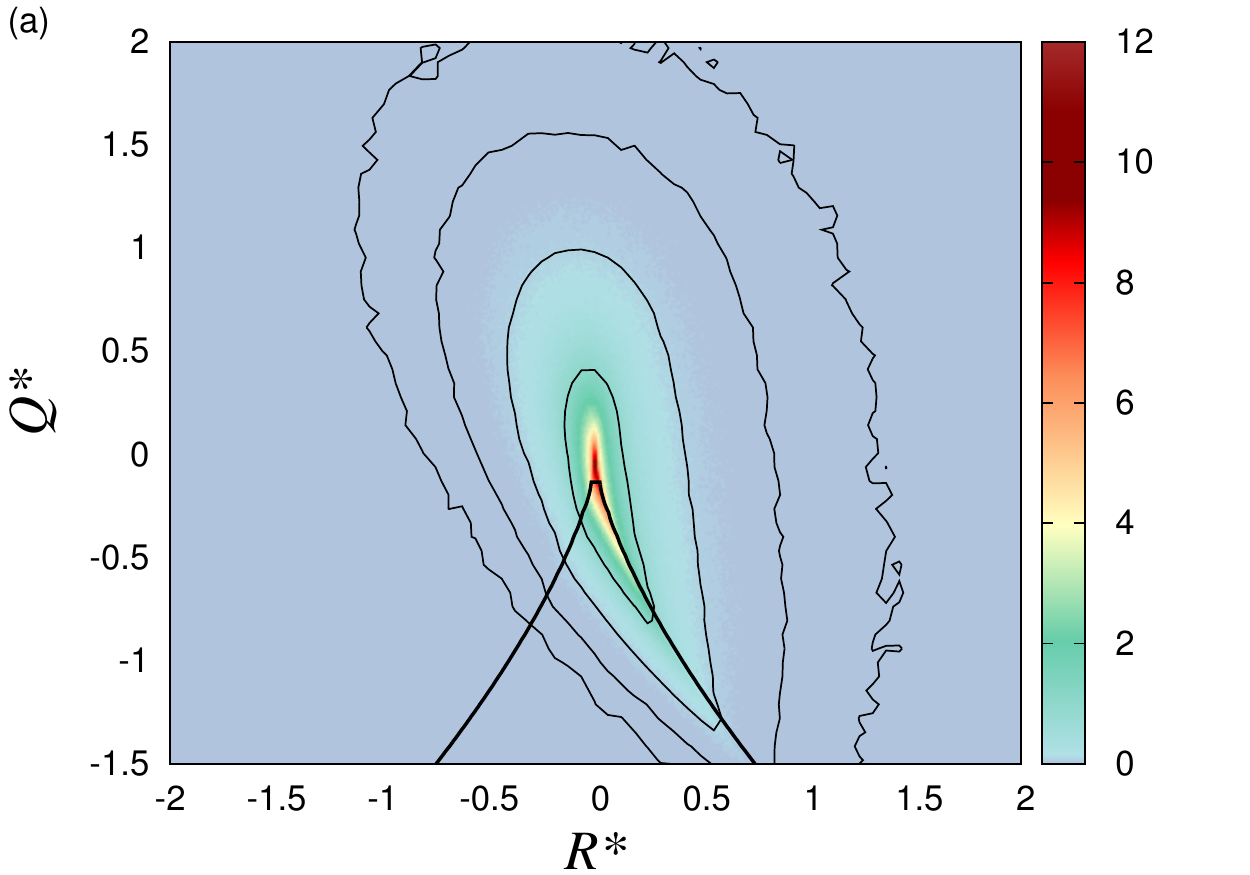}
\includegraphics[width=7.8cm, height=5.39cm]{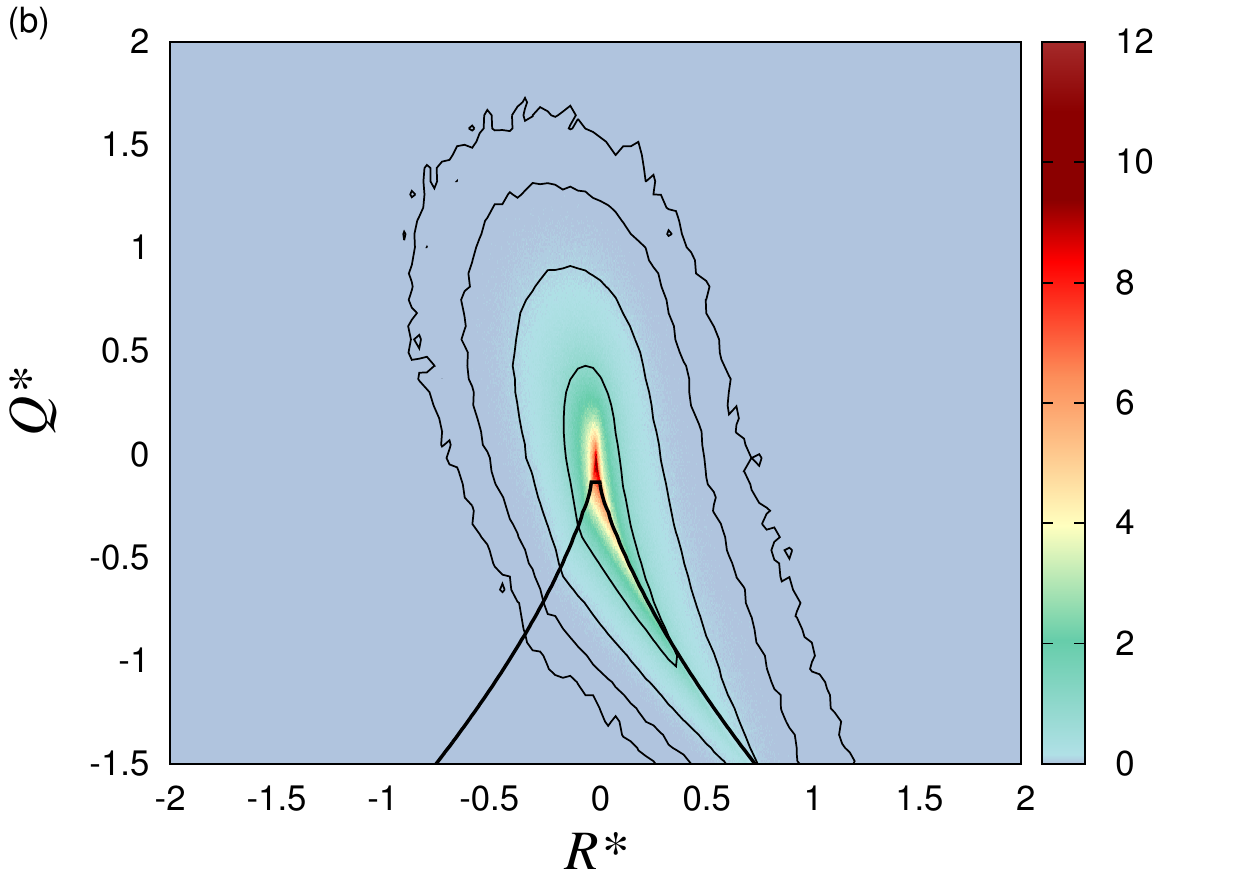}
\label{}
\vspace{-0.0cm}

\caption{Contour plots for the joint PDFs of
the normalized Cayley-Hamilton invariants $Q^\ast$ and $R^\ast$ for $\tau=0.1$
and $g=\sqrt{2}$. The level curves have PDF values equal to 1, $10^{-1}$, $10^{-2}$
and $10^{-3}$. Figure (a) is obtained from the the direct numerical 
integration of the RFDC stochastic equations, while figure (b) is evaluated 
from the analytical vgPDF discussed in Sec.~III. The inverted V-shaped lines 
in both figures (a) and (b) indicate the Vieillefosse zero discriminant line.}
\end{figure}
\end{center} 
\newpage

Samples of non-diagonal and diagonal components of the velocity gradient have been grouped into two distinct sets. We refer, thus, to vgPDfs of non-diagonal and diagonal components of $\A$, without specifying any particular cartesian tensor indices. Our statistical evaluations have been performed with sets of $12 \times 10^6$ and $24 \times 10^6$ elements for the diagonal and non-diagonal components, respectively, of the velocity gradient tensor.

Our results are shown in Figs.~2-4. In order to appreciate the relevance of noise renormalization, we also have depicted how would the vgPDFS look like if the noise vertex were not corrected by the loop diagram of Fig.~1 (these PDFs are given in Figs.~2a, 2c, 3a, and 3c). The one-loop correction leads, in fact, to much better approximations for the vgPDFs. The results are even more satisfactory to the eye if the vgPDFs are plotted in linear scales (Fig.~3), since larger deviations from the analytical expressions are found mostly in the far tails of the vgPDFs and are actually associated to small cumulative probabilities.

In Fig.~4 we show how the analytical and the empirical joint probability distributions of the normalized Cayley-Hamilton invariants 
\be
Q^\ast = -\frac{\tr(\A^2)}{2 \langle S^2 \rangle} {\hbox{ and }} R^\ast = - \frac{\tr (\A^3)}{3 \langle S^2 \rangle^{3/2}} \ , \
\ee
where $S$ is the rate-of-strain tensor, with $S^2 \equiv S_{ij}S_{ij}$, compare to each other. We find that the analytical joint PDF is able to account for the essential qualitative geometrical features as the ``tear-drop" shapes of the level curves and the role of the zero-discriminant line. The quantitative agreement is better, of course, close to the origin of the $(R^\ast,Q^\ast)$ plane, where non-linear fluctuations of the velocity gradient tensor tend to be suppressed.

\section{\leftline{Conclusions}}

We have carried out an analytical study of the vgPDFs in the RFDC lagrangian model of turbulence.
The MSR framework in its path-integral formulation proves to be a very convenient setup, where standard
field-theoretical semiclassical approaches can be straightforwardly applied. Once it is difficult to establish 
exact saddle-point solutions for the Euler-Lagrange equations associated either to the bare or to the effective 
MSR action, we have used, as an approximation, solutions that hold in the regime of small noise strength. A 
further source of technical difficulty is related to the precise evaluation of the effective MSR action up to 
one-loop order: in fact, one should take into account a large number of vertex corrections, leading to non-local 
kernels as the result of much more involved computations. We have, thus, put forward a pragmatical strategy 
for the evaluation of the effective MSR action where, as working hypotheses, (i) only the noise vertex is 
corrected up to one-loop order and (ii) a low-frequency approximation for the renormalized noise vertex is 
implemented. Nevertheless the above simplifying assumptions, the resulting analytical vgPDFs are satisfactorily 
compared to the empirical ones for a meaningful range of bare noise coupling constants ($g < \sqrt{2}$). We leave 
for additional research the necessary refinements on the approach we have adopted in this 
paper. We also note that time-dependent correlation functions of the velocity gradient tensor can be 
evaluated along similar semiclassical lines.

Analytical vgPDFs are a promising tool in the study of turbulent intermittency. Once validated, 
they can be used to investigate conditional statistics phenomena in a way that would be not possible through 
the ensembles produced from solutions of the related stochastic differential equations. A particularly interesting 
application of the analytical vgPDFS can be attempted, in principle, in the context of turbulent geometrical 
statistics, in order to clarify the statistical relations between the vorticity 
and the rate-of-strain fields.

It is important to emphasize, as a concluding remark, that the MSR semiclassical method as discussed 
in this work can be straightforwardly applied, with no further conceptual or technical obstacles, to 
several turbulence models and to a large class of phase-space reduced stochastic dynamical systems. 

\vspace{0.4cm}

{\leftline{\small{\bf{ACKNOWLEDEGMENTS}}}}
\vspace{0.4cm}

This work has been partially supported by CNPq and FAPERJ. The authors have
greatly benefited from the use of NIDF facilities at COPPE-UFRJ. One of the 
authors (L.M.) would like to thank the warm hospitality of the ICTP during 
early stages of this work.

\end{document}